\documentclass[12pt]{article}
\usepackage{amsmath}
\usepackage{amscd}
\usepackage{amssymb}
\usepackage{array}

\newcommand{\R}{\mathbb{R}}
\newcommand{\C}{\mathbb{C}}
\newcommand{\Z}{\mathbb{Z}}

\newcommand{\A}{\mathcal{A}}

\newcommand{\rO}{\mathrm{O}}
\newcommand{\rSO}{\mathrm{SO}}

\begin{document}
\rightline{CERN-TH/2003-150}

\rightline{IFIC/03-32}

\rightline{FTUV-03/0602}

 \vskip 2.5cm

  \centerline{\Large \bf Supersymmetry in noncommutative superspaces}

\vskip 2cm

\centerline{S. Ferrara$^\flat$$^\dag$, M. A. Lled\'o$^\sharp$ and
O. Maci\'a$^\natural$}
\bigskip
\bigskip

\centerline{\it $^\flat$ CERN, Theory Division, CH 1211 Geneva 23,
Switzerland.}

\bigskip

\centerline{\it $^\sharp$ INFN, Sezione di Torino and}
\centerline{\it
 Dipartimento di Fisica, Politecnico di Torino,}
\centerline{\it Corso Duca degli Abruzzi 24, I-10129 Torino,
Italy.}

\bigskip

\centerline{\it $^\natural$ Departament de F\'{\i}sica  Te\`orica,
Universitat de Val\`encia,}
 \centerline{\it C. Dr.
Moliner, 50 E-46100, Burjassot. Val\`encia, Spain.}

\bigskip
\centerline{\it $^\dag$ INFN, Laboratori Nazionali di Frascati,}
\centerline{\it Via E. Fermi,40. I-00044 Frascati (Rome) Italy.}

\vskip 2cm
\begin{abstract}
  Non commutative superspaces can be introduced as the Moyal-Weyl
  quantization of a Poisson bracket for classical superfields.
  Different deformations are studied corresponding to constant
  background fields in string theory. Supersymmetric and non
  supersymmetric deformations can be defined, depending on the
  differential operators used to define the Poisson bracket. Some
  examples of deformed, 4 dimensional lagrangians are given. For
  extended superspace ($N>1$), some new deformations can be
  defined, with no analogue in the $N=1$ case.

\end{abstract}

\vfill\eject
\section{Introduction}

Non commutative geometry  and supergeometry naturally arise in
string theory in several contexts. Among others, we may mention
the work of Connes, Douglas and Schwarz \cite{cds} where non
commutative tori where introduced as possible compactification
spaces of M theory, the work of  Banks, Fischler, Shenker and
Susskind \cite{bfss}
 where M-theory was related to  the $N \rightarrow\infty$ limit
 of the supersymmetric matrix
quantum mechanics describing D0-branes and the work of Seiberg and
Witten \cite{sw} where a certain limit of the string dynamics is
described by a gauge theory in presence of a non zero background
field $B_{\mu\nu}$. (For a review, see Ref. \cite{dn}, and
references therein).

More recently, the extension of non commutativity to odd variables
has been related to the presence of other background fields. The
R-R field strength backgrounds  give rise to a deformation of type
$\theta-\theta$ \cite{ov}, and the gravitino background gives rise
to an $x-\theta$ deformation \cite{bgn}.

Field theories in non commutative spaces have been considered in
the literature in a broader context, where a huge amount of work
has been done. For an introduction to different aspects of the
subject, see for example Refs. \cite{dfr,co,la,mssw,bcz,av}.

In connection with string theory, supersymmetric theories have
been considered mainly where the deformation of superspace affects
only to the spacetime part ($x-x$ deformations). There, the
extension of Wess-Zumino \cite{cz,fl} and Yang-Mills models is
straightforward  once the gauge invariance in superspace is
properly defined \cite{fl}. Also in Ref. \cite{fl} the possibility
of a supersymmetric deformation of superspace with non vanishing
$\theta-\theta$ deformation was explored. In fact, the possibility
of having fermi coordinates that have a non zero anticommutator
was already studied in the literature \cite{bs2,sn}. From a more
mathematical point of view, non commutative supermanifolds and
supervarieties have also been considered in the literature
\cite{sc,fl2,hiu}.

Starting from the observation of Ooguri and Vafa \cite{ov},
deformations of the anticommuting variables have acquired renewed
interest. The effect of such deformations in the Lagrangian has
been investigated by Seiberg \cite{se} and  Berkovits and Seiberg
\cite{bs}.

\bigskip

In the present paper we consider a variety of deformations, both
for $N=1$ and extended ($N=2$) supesymmetry in $D=4$. These
deformations vary according the differential operators chosen to
construct the Poisson bracket that afterwards becomes quantized
with a star product of Moyal-Weyl type. In particular, we show
explicitly the difference between the deformation considered in
Ref. \cite{fl} and the one proposed in Ref. \cite{se}.  The first
one has the advantage of being manifestly supersymmetric, the
second one, although it explicitly breaks one half of the
supersymemtry, allows the definition of chiral and antichiral
superfields, which form subalgebras of the star product.  This was
not possible with the supersymmetric deformation, and allows a
simple generalization of super Yang-Mills theories to the deformed
superspace. For the Wess-Zumino model both deformations lead to
the same Lagrangian, preserving 1/2 of the supersymmetry.

We also consider deformations of type $x-\theta$ and explore the
consequences of these in the Lagrangian for some simple cases.

Finally we study deformations of extended superspaces where new
possibilities arise. For example, one can consistently have a non
trivial $\theta-\bar\theta$ anticommutator, which is related to
having constant vector backgrounds. This corresponds to
deformations of harmonic superspace \cite{gios}.

\bigskip

The paper is organized as follows. In section \ref{generalities}
we remind some known facts about Poisson brackets in superspace
and fix the notation. Section \ref{supersymmetric} is devoted to
the definition of supersymmetric Poisson brackets and their
Moyal-Weyl deformation. In particular, we obtain the formula for
the Moyal-Weyl deformation in presence of an $x-\theta$
deformation. In section \ref{extended} we consider extended
superspace and its deformations. In section \ref{wz} we study the
deformed Wess-Zumino model in different scenarios.

\section{\label{generalities}Generalities on super Poisson brackets.}
We consider the superalgebra $\A= C^\infty(\R^n)\otimes
\Lambda(\R^m)$ which has an obvious $\Z_2$ grading. We say that an
even element $\phi$ has parity $p(\phi)=p_\phi=0$ and an odd
element has parity $p(\phi)=p_\phi=1$. We have that
$$p(\phi\psi)=p(\phi)+p(\psi)\;\; \mathrm{mod} (2)$$
Even and odd elements are homogeneous elements.

A Poisson bracket on $\A$ is a bilinear operation
$$\{\;,\;\}:\A\times\A\rightarrow \A$$
such that for homogeneous elements $\phi, \psi$ and $\chi$

\noindent 1. $p(\{\phi,\psi\})= p(\phi)+p(\psi)$ (it is an {\it
even} Poisson bracket).

\noindent 2. $\{\phi,\psi\}=-(-1)^{p_\phi p_\psi}\{\psi,\phi\}$.

\noindent 3.(Derivation property)
\begin{eqnarray*}\{\phi\phi',\psi\}=\phi\{\phi',\psi\}+(-1)^{p_\phi'
p_\psi}\{\phi,\psi\}\phi'\\
\{\phi,\psi\psi'\}= (-1)^{p_\psi p_\phi}\phi\{\phi,\psi'\}+
\{\phi,\psi\}\psi'
\end{eqnarray*}

\noindent 4. (Graded Jacobi identity)
$$\{\phi,\{\psi,\chi\}\}+ (-1)^{p_\chi(p_\phi+p_\psi)}\{\chi,\{\phi,\psi\}\}
+ (-1)^{p_\phi(p_\psi+p_\chi)}\{\psi,\{\chi,\phi\}\}=0.$$

\bigskip

Let us  denote \begin{eqnarray*}&A=a, \qquad &\hbox{for} \quad A=1,\dots n,\\
&A=\alpha +n, \qquad &\hbox{for} \quad A=n+1,\dots
n+m.\end{eqnarray*} $z^A$ denotes generically all variables $x^a$
and $\theta^\alpha$. Then, a Poisson bracket on $\A$ can be
written as
\begin{equation}\{\phi,\psi\}=\phi\overleftarrow{\partial}_A
P^{AB}\overrightarrow{\partial}_B\psi, \label{spb}\end{equation}
where
$$\overleftarrow{\partial}_A= \frac {\overleftarrow{\partial}}{\partial z^A},\qquad
\overrightarrow{\partial}_A= \frac
{\overrightarrow{\partial}}{\partial z^A}$$ are respectively right
and left derivatives. They both coincide with the ordinary
derivative in the case that $z^A$ is an even variable. For odd
variables one has \begin{eqnarray*}(f(x)\theta^{\alpha_1}\dots
\theta^{\alpha_k}\xi \theta^{\beta_1}\dots
\theta^{\beta_l})\overleftarrow{\partial}_\xi=(-1)^l
(f(x)\theta^{\alpha_1}\dots \theta^{\alpha_k}
\theta^{\beta_1}\dots
\theta^{\beta_l}), \\
\overrightarrow{\partial}_\xi(f(x)\theta^{\alpha_1}\dots
\theta^{\alpha_k}\xi \theta^{\beta_1}\dots
\theta^{\beta_l})=(-1)^k (f(x)\theta^{\alpha_1}\dots
\theta^{\alpha_k} \theta^{\beta_1}\dots
\theta^{\beta_l}),\end{eqnarray*}so $$
\phi\overleftarrow{\partial}_A=
(-1)^{p_A(p_\phi+1)}\overrightarrow{\partial}_A\phi .$$ They are
odd derivations of the algebra $\A$:
\begin{eqnarray*}&\overrightarrow{\partial}_\xi (\phi\psi)=
\overrightarrow{\partial}_\xi (\phi)\psi+(-1)^{p_\xi
p_\phi}\phi\overrightarrow{\partial}_\xi (\psi), \qquad &\hbox{(left derivation)}, \\
&(\phi\psi)\overleftarrow{\partial}_\xi =
\phi(\psi)\overleftarrow{\partial}_\xi +(-1)^{p_\xi
p_\psi}(\phi)\overleftarrow{\partial}_\xi \psi,\qquad
&\hbox{(right derivation)}.
\end{eqnarray*}
Right and left derivatives are chosen in such way that for $F\in
\A$
$$dF=F\overleftarrow{\partial}_\xi
d\theta=d\theta\overrightarrow{\partial}_\xi F.$$

 If
$P^{AB}=-(-1)^{p_Ap_B}P^{BA}$ and $ p(P^{AB})=p_A+p_B,$ then
properties 1,2,3 are automatically satisfied.

If we want to consider  a constant Poisson bracket
($P^{AB}=\mathrm{constant}$, that is, independent of $z^A$), we
have to extend the scalars to a commutative superalgebra. Then the
entries of $R$ and $S$ are {\it odd scalars}. The algebra will be
$\A[\xi^1,\dots \xi^s]=\C^\infty(\R^n)\otimes \Lambda(\R^m)\otimes
\Lambda(\R^s)$, where the odd generators $\xi^i \in \Lambda(\R^s)$
are considered as scalars (inert under the derivations). Then, the
graded Jacobi identity is also satisfied.

\section{\label{supersymmetric}Supersymmetric Poisson brackets and star products}
  We consider a four dimensional space time with
coordinates $x^\mu$, $\mu=0,\dots 3$ and  Minkowskian signature.
For the moment being we take $N=1$ supersymmetry, with one complex
Weyl spinor $\theta^\alpha$, $\alpha=1,2$, generating the odd part
of the superspace.  On the space of superfields the covariant left
derivatives are defined as
\begin{eqnarray*}
\overrightarrow{D}_\alpha\Phi & =
&\overrightarrow{\partial}_\alpha\Phi
 +i
\sigma^\mu_{\alpha\dot\alpha}\bar\theta^{\dot\alpha}\partial_\mu\Phi\\
\overrightarrow{\bar D}_{\dot\alpha}\Phi & =&
-\overrightarrow{\partial}_{\dot\alpha}\Phi
 -i
\theta^\alpha\sigma^\mu_{\alpha\dot\alpha}\partial_\mu\Phi
\end{eqnarray*} satisfying an algebra
\begin{eqnarray*}
\{\overrightarrow{D}_\alpha,\overrightarrow{D}_\beta\}=
\{\overrightarrow{\bar D}_{\dot \alpha},\overrightarrow{\bar D}_{\dot\beta}\}=0\\
\{ \overrightarrow{D}_{\alpha},\overrightarrow{\bar
D}_{\dot\alpha}\}=-2i\sigma^\mu_{\alpha\dot\alpha}{\partial_\mu}.
\end{eqnarray*}
Given a left derivation $\overrightarrow{\mathcal{D}}$ of degree
$p_\mathcal{D}$, that is,
$$\overrightarrow{\mathcal{D}}(\Phi\Psi)=\overrightarrow{\mathcal{D}}(\Phi)\Psi +
(-1)^{p_\mathcal{D}p_\Phi}\Phi\overrightarrow{\mathcal{D}}(\Psi),$$
one can define a right  derivation $\overleftarrow{\mathcal{D}}$,
also  of degree $p_\mathcal{D}$
$$(\Phi\Psi)\overleftarrow{\mathcal{D}}=\Phi(\Psi)\overleftarrow{\mathcal{D}}+ (-1)^{p_\mathcal{D}p_\Psi}
(\Phi)\overleftarrow{\mathcal{D}}\Psi,$$ in the following way
$$\Phi\overleftarrow{\mathcal{D}}=
(-1)^{p_\mathcal{D}(p_\Phi+1)}\overrightarrow{\mathcal{D}}\Phi.$$
Then we can define the right covariant derivatives as
\begin{eqnarray*}
\Phi \overleftarrow{D}_\alpha &=& \Phi
\overleftarrow{\partial}_\alpha -i\partial_\mu\Phi
\sigma^\mu_{\alpha\dot\alpha}\bar\theta^{\dot\alpha}\\
\Phi \overleftarrow{\bar D}_{\dot\alpha} &=& -\Phi
\overleftarrow{\bar\partial}_{\dot\alpha} +i\partial_\mu\Phi
\theta^\alpha\sigma^\mu_{\alpha\dot\alpha}. \end{eqnarray*}
 They satisfy an algebra
 \begin{eqnarray*}
\{\overleftarrow{D}_\alpha,\overleftarrow{D}_\beta\}=
\{\overleftarrow{\bar D}_{\dot \alpha},\overleftarrow{\bar D}_{\dot\beta}\}=0\\
\{ \overleftarrow{D}_{\alpha},\overleftarrow{\bar
D}_{\dot\alpha}\}=+2i\sigma^\mu_{\alpha\dot\alpha}{\partial_\mu}.
\end{eqnarray*}
On may notice that for any two odd derivations
\begin{equation}[\overleftarrow{D},\overrightarrow{D'}]\Phi=(-1)^{p_\Phi
}\{\overrightarrow{D},\overrightarrow{D'}\}\Phi.
\label{comder}\end{equation} The supersymmetry algebra is realized
as an algebra of left derivations on superspace
\begin{eqnarray*}
\overrightarrow{Q}_\alpha  = \overrightarrow{\partial}_\alpha
 -i
\sigma^\mu_{\alpha\dot\alpha}\bar\theta^{\dot\alpha}\partial_\mu,
\qquad  {\overrightarrow{\bar Q}}_\alpha  =
-\overrightarrow{\bar\partial}_{\dot\alpha}
 +i
\theta^\alpha\sigma^\mu_{\alpha\dot\alpha}\partial_\mu,
\end{eqnarray*}
satisfying
\begin{eqnarray*}
&&\{\overrightarrow{Q}_\alpha,\overrightarrow{Q}_\beta\}=\{\overrightarrow{\bar
Q}_{\dot \alpha},
\overrightarrow{\bar Q}_{\dot\beta}\}=0\\
&&\{ \overrightarrow{Q}_{\alpha},\overrightarrow{\bar
Q}_{\dot\alpha}\}=+2i\sigma^\mu_{\alpha\dot\alpha}{\partial_\mu}.
\end{eqnarray*}
One has also
\begin{equation}
\{\overrightarrow{D}_\alpha,\overrightarrow{Q}_\beta\}=\{\overrightarrow{D}_\alpha,
\overrightarrow{\bar Q}_{\dot \beta}\}=\{\overrightarrow{\bar
D}_{\dot \alpha},\overrightarrow{Q}_\beta\}=
\{\overrightarrow{\bar D}_{\dot\alpha},\overrightarrow{\bar
Q}_{\dot\beta}\}=0.\label{DQcr}\end{equation}
\subsection{Supersymmetric Poisson brackets}

 On the space of complex superfields one can consider
the following Poisson bracket, which is a generalization of the
Poisson bracket considered in  \cite{fl}:
\begin{equation}\{\Phi, \Psi\}_1=P^{\mu\nu}\frac{\partial
\Phi}{\partial x^\mu}\frac{\partial \Psi}{\partial x^\nu}+
P^{\alpha\beta}\Phi\overleftarrow{D}_\alpha\overrightarrow{D}_\beta\Psi
+ \frac{\partial \Phi}{\partial x^\mu}
P^{\mu\alpha}\overrightarrow{D}_\alpha\Psi-
\Phi\overleftarrow{D}_\alpha P^{\mu\alpha} \frac{\partial
\Psi}{\partial x^\mu} .\label{susypb1}\end{equation} This
satisfies the Jacobi identity with $P^{\mu\nu}$ antisymmetric,
$P^{\alpha\beta}$ symmetric, $P^{\mu\alpha}$ arbitrary (and odd), and all of
them constant. It will be convenient to take $P^{\mu\nu}$ pure
imaginary, while the other matrices are just complex. Because of
(\ref{comder}) and (\ref{DQcr}), $\{\;,\;\}_1$ is a {\it
supersymmetric} Poisson bracket, that is, the supersymmetry
charges are derivations with respect to the Poisson bracket,
\begin{eqnarray*}\overrightarrow{Q}\{\Phi, \Psi\}=\{\overrightarrow{Q}\Phi,
\Psi\}+(-1)^{p_\Phi}\{\Phi,
\overrightarrow{Q}\Psi\}\\\overrightarrow{\bar Q}\{\Phi,
\Psi\}=\{\overrightarrow{\bar Q}\Phi, \Psi\}+(-1)^{p_\Phi}\{\Phi,
\overrightarrow{\bar Q}\Psi\} .\end{eqnarray*} One can replace $D$
by $\bar D$ in (\ref{susypb1}) and write another Poisson bracket
\begin{equation*}\{\Phi,
\Psi\}_2=P^{\mu\nu}\frac{\partial \Phi}{\partial
x^\mu}\frac{\partial \Psi}{\partial x^\nu}+
P^{\dot\alpha\dot\beta}\Phi\overleftarrow{\bar
D}_{\dot\alpha}\overrightarrow{\bar D}_{\dot\beta}\Psi+
\frac{\partial \Phi}{\partial x^\mu}
P^{\mu\dot\alpha}\overrightarrow{\bar D}_{\dot\alpha}\Psi-
\Phi\overleftarrow{\bar
D}_{\dot\alpha}P^{\mu\dot\alpha}\frac{\partial \Psi}{\partial
x^\mu}.\end{equation*} but one cannot have in principle terms
with both derivatives, $D$ and $\bar D$, since they do not anticommute,
and the Jacobi identity will not be immediately satisfied. A consequence of
this is that $\{\;,\;\}_{1,2}$ are degenerate in the space of odd
variables. Another consequence is that in Minkowskian space the
Poisson brackets do not satisfy a reality condition,
$$\{\Phi^*,\Psi^*\}_{1,2}\neq \{\Psi,
\Phi\}_{1,2}^*,$$ which does not mean that they are ill defined
nor inconsistent in any way. If we take
$(P^{\alpha\beta})^*=P^{\dot\alpha\dot\beta}$ and
$(P^{\mu\alpha})^*=-P^{\mu\dot\alpha}$, they satisfy instead the
relation
$$\{\Phi^*,\Psi^*\}_{1,2}= \{\Psi,
\Phi\}_{2,1}^*.$$ In Section \ref{ssdwzm} we will use the
quantization of these Poisson brackets to give a prescription for
a deformed Wess-Zumino Lagrangian which is invariant with respect
to the whole supertranslation algebra.

In other signatures or dimensions, a similar, supersymmetric
Poisson bracket may admit such a reality condition \cite{kpt}.

Chiral superfields, $\bar D_{\dot \alpha}\Phi=0$, are a Poisson
subalgebra of  $\{\;,\;\}_{2}$ (as antichiral superfields,
$D_{\dot \alpha}\Phi=0$, are a Poisson subalgebra of
$\{\;,\;\}_{1}$). Indeed, the Poisson structures restricted to
these subspaces involve only the even coordinates $x^\mu$.

It is useful to express $\{\;,\;\}_{1,2}$ in terms of ordinary
derivatives:

\begin{eqnarray*}
&&\{\Phi, \Psi\}_1=\bigl(P^{\mu\nu}+C^{\mu\nu}\bar\theta\bar\theta
+iP^{\mu\alpha}\sigma^\nu_{\alpha\dot\alpha}\bar\theta^{\dot\alpha}-
iP^{\nu\alpha}\sigma^\mu_{\alpha\dot\alpha}\bar\theta^{\dot\alpha}
\bigr)\partial_\mu\Phi\partial_\nu\Psi\\&
&+\partial_\mu\Phi\bigl(iP^{\alpha\beta}\sigma^\mu_{\alpha\dot\alpha}\bar\theta^{\dot\alpha}+P^{\mu\beta}\bigr)
\overrightarrow{\partial}_\beta\Psi-
\Phi\overleftarrow{\partial}_\alpha \bigl(
iP^{\alpha\beta}\sigma^\nu_{\beta\dot\beta}\bar\theta^{\dot \beta}
+P^{\nu\alpha}\bigr)\partial_\nu\Psi\\&&+
P^{\alpha\beta}\Phi\overleftarrow{\partial}_\alpha\overrightarrow{\partial}_\beta\Psi,\\\\&&
\{\Phi, \Psi\}_2=\bigl(P^{\mu\nu}+\theta \theta
D^{\mu\nu}-iP^{\mu\dot\alpha}\theta^\alpha\sigma^\nu_{\alpha\dot\alpha}
+iP^{\mu\dot\alpha}\theta^\alpha\sigma^\mu_{\alpha\dot\alpha}
\bigr)\partial_\mu\Phi\partial_\nu\Psi\\&&
+\partial_\mu\Phi\bigl(iP^{\dot\alpha\dot\beta}\theta^{\alpha}\sigma^\mu_{\alpha\dot\alpha}-P^{\mu\dot\beta}\bigr)
\overrightarrow{\bar\partial}_{\dot\beta}\Psi-
\Phi\overleftarrow{\bar\partial}_{\dot\alpha}\bigl(iP^{\dot\alpha\dot\beta}\theta^{\beta}\sigma^\nu_{\beta\dot\beta}
-P^{\nu\dot\alpha}\bigr)\partial_\nu\Psi\\&&+
P^{\dot\alpha\dot\beta}\Phi\overleftarrow{\bar\partial}_{\dot\alpha}\overrightarrow{\bar\partial}_{\dot\beta}\Psi,\end{eqnarray*}
where
$$C^{\mu\nu}= \frac 12
P^{\alpha\beta}\sigma^\mu_{\alpha\dot\alpha}
\sigma^\nu_{\beta\dot\beta}\epsilon^{\dot \alpha\dot\beta}, \qquad
D^{\mu\nu}=-\frac12P^{\dot\alpha\dot\beta}\epsilon^{\alpha\beta}\sigma^\mu_{\alpha\dot\alpha}
\sigma^\nu_{\beta\dot\beta}.
$$ From these formulas  the commutation rules of the basic
variables $x^\mu, \theta^\alpha, \bar \theta^{\dot\alpha}$ can be
read directly.

We make now a change of variables \cite{fzw}
$$x^\mu, \; \theta^\alpha,\; \bar\theta^{\dot\alpha}\;\longrightarrow \; y^\mu=x^\mu +i\theta^\alpha
\sigma^\mu_{\alpha\dot\alpha} \bar\theta^{\dot \alpha},\;
\theta^\alpha,\;\bar\theta^{\dot\alpha}.$$ A superfield may be
expressed in both coordinate systems
$$\Phi(x, \theta, \bar\theta)=\Phi'(y, \theta, \bar\theta).$$ The
covariant derivatives and supersymmetry charges take the form
\begin{eqnarray*}
\overrightarrow{D}_\alpha\Phi'  =
\overrightarrow{\partial}_\alpha\Phi'
 +2i
\sigma^\mu_{\alpha\dot\alpha}\bar\theta^{\dot\alpha}
\frac{\partial \Phi' }{\partial y^\mu}\qquad \overrightarrow{\bar
D}_{\dot\alpha}\Phi' =
-\overrightarrow{\partial}_{\dot\alpha}\Phi',\\
  {\overrightarrow{\bar Q}}_{\dot\alpha}\Phi'  =
-\overrightarrow{\partial}_{\dot\alpha}\Phi' +2i
\theta^\alpha\sigma^\mu_{\alpha\dot\alpha}\frac{\partial \Phi'
}{\partial y^\mu}\qquad  \overrightarrow{Q}_\alpha\Phi' =
\overrightarrow{\partial}_\alpha\Phi'.
\end{eqnarray*}

In the new coordinates the brackets $\{\;,\;\}_{1,2}$ become
\begin{eqnarray*}
&&\{\Phi',
\Psi'\}_1=\bigl(P^{\mu\nu}+4C^{\mu\nu}\bar\theta\bar\theta
+2iP^{\mu\alpha}\sigma^\nu_{\alpha\dot\alpha}\bar\theta^{\dot\alpha}-
2iP^{\nu\alpha}\sigma^\mu_{\alpha\dot\alpha}\bar\theta^{\dot\alpha}
\bigr)\frac{\partial\Phi'}{\partial
y^\mu}\frac{\partial\Psi'}{\partial y^\nu}\\&& +\frac{\partial
\Phi'}{\partial
y^\mu}\bigl(2iP^{\alpha\beta}\sigma^\mu_{\alpha\dot\alpha}\bar\theta^{\dot\alpha}+P^{\mu\beta}\bigr)
\overrightarrow{\partial}_\beta\Psi'-
\Phi'\overleftarrow{\partial}_\alpha \bigl(
2iP^{\alpha\beta}\sigma^\nu_{\beta\dot\beta}\bar\theta^{\dot
\beta} +P^{\nu\alpha}\bigr)\frac{\partial \Psi'}{\partial
y^\nu}\\&&+
P^{\alpha\beta}\Phi'\overleftarrow{\partial}_\alpha\overrightarrow{\partial}_\beta\Psi',\\\\&&\{\Phi',
\Psi'\}_2=P^{\mu\nu}\frac{\partial \Phi'}{\partial
y^\mu}\frac{\partial \Psi'}{\partial y^\nu}+
P^{\dot\alpha\dot\beta}\Phi'\overleftarrow{\bar
\partial}_{\dot\alpha}\overrightarrow{\bar
\partial}_{\dot\beta}\Psi'\\&&+ \frac{\partial \Phi'}{\partial y^\mu}
P^{\mu\dot\alpha}\overrightarrow{\bar
\partial}_{\dot\alpha}\Psi'- \Phi'\overleftarrow{\bar
\partial}_{\dot\alpha}P^{\mu\dot\alpha}\frac{\partial
\Psi'}{\partial y^\mu},\end{eqnarray*}
 which simplifies $\{\;,\;\}_{2}$.

\subsection{\label{nonsusy}Non supersymmetric Poisson brackets}

One can define different Poisson brackets by making use of the
operators $Q_\alpha$ and $\bar Q_{\dot \alpha}$. In fact, the
operators $D$'s and the $Q$'s play interchangeable roles. Consider
for example the brackets \cite{se}
\begin{eqnarray}\{\Phi,
\Psi\}_3=P^{\mu\nu}\frac{\partial \Phi}{\partial
x^\mu}\frac{\partial \Psi}{\partial x^\nu}+
P^{\alpha\beta}\Phi\overleftarrow{Q}_\alpha\overrightarrow{Q}_\beta\Psi
+ \frac{\partial \Phi}{\partial x^\mu}
P^{\mu\alpha}\overrightarrow{Q}_\alpha\Psi-
\Phi\overleftarrow{Q}_\alpha P^{\mu\alpha} \frac{\partial
\Psi}{\partial x^\mu}
\label{suse}\\\nonumber\\
\{\Phi, \Psi\}_4=P^{\mu\nu}\frac{\partial \Phi}{\partial
x^\mu}\frac{\partial\Psi}{\partial x^\nu}+
P^{\dot\alpha\dot\beta}\Phi\overleftarrow{\bar
Q}_{\dot\alpha}\overrightarrow{\bar Q}_{\dot\beta}\Psi +
\frac{\partial \Phi}{\partial x^\mu}
P^{\mu\dot\alpha}\overrightarrow{\bar
Q}_{\dot\alpha}\Psi-\Phi\overleftarrow{\bar
Q}_{\dot\alpha}P^{\mu\dot\alpha}\frac{\partial \Psi}{\partial
x^\mu} ,\nonumber\end{eqnarray} where the right acting charges are
\begin{eqnarray*}
\Phi \overleftarrow{Q}_\alpha &=& \Phi
\overleftarrow{\partial}_\alpha +i\frac{\partial \Phi}{\partial
x^\mu}
\sigma^\mu_{\alpha\dot\alpha}\bar\theta^{\dot\alpha}\\
\Phi \overleftarrow{\bar Q}_{\dot\alpha} &=& \Phi
\overleftarrow{\bar\partial}_{\dot\alpha} +i\frac{\partial
\Psi}{\partial x^\mu} \theta^\alpha\sigma^\mu_{\alpha\dot\alpha}.
\end{eqnarray*}

In terms of the coordinates $y,\theta, \bar\theta$, the bracket
$\{\;,\;\}_3$, for example, becomes
$$\{\Phi',
\Psi'\}_3=P^{\mu\nu}\frac{\partial \Phi'}{\partial
y^\mu}\frac{\partial \Psi'}{\partial y^\nu}+
P^{\alpha\beta}\Phi'\overleftarrow{\partial}_\alpha\overrightarrow{\partial}_\beta\Psi'+
\frac{\partial \Phi}{\partial y^\mu} P^{\mu\alpha}
\overrightarrow{\partial}_\alpha\Psi- \Phi\overleftarrow{\partial
}_\alpha P^{\mu\alpha}\frac{\partial \Psi}{\partial y^\mu}.$$ The
quantization of this bracket with $P^{\mu\nu}=P^{\mu\alpha}=0$ was
explored in Ref. \cite{se}. Since $\{Q,\bar Q\}\neq 0$, this
bracket is not supersymmetric with respect to the charges $\bar
Q$, that is,
$$\overrightarrow{\bar Q}\{\Phi, \Psi\}_3\neq \{\overrightarrow{\bar Q}\Phi,
 \Psi\}_3+(-1)^{p_\Phi}\{\Phi, \overrightarrow{\bar Q}\Psi\}_3,$$
 although it is still supersymmetric with respect to the charges
 $Q$. Because of (\ref{DQcr}), one has instead that the operators
 $D$ and $\bar D$ are derivations with respect to this bracket,
 \begin {eqnarray*}
 \overrightarrow{ D}\{\Phi, \Psi\}_{3,4}= \{\overrightarrow{ D}\Phi,
 \Psi\}_{3,4}+(-1)^{p_\Phi}\{\Phi, \overrightarrow{ D}\Psi\}_{3,4}\\
 \overrightarrow{\bar D}\{\Phi, \Psi\}_{3,4}= \{\overrightarrow{\bar D}\Phi,
 \Psi\}_{3,4}+(-1)^{p_\Phi}\{\Phi, \overrightarrow{\bar
 D}\Psi\}_{3,4}.\end{eqnarray*}
 It follows that the subspaces of chiral ($\bar D\Phi$=0) and antichiral ($
 D\Phi$=0) superfields are Poisson subalgebras of $\{\;,\;\}_{3,4}$.

 \bigskip

 One could use a different change of variables
 $$x^\mu, \; \theta^\alpha,\; \bar\theta^{\dot\alpha}\;\longrightarrow \; \bar y^\mu=x^\mu -i\theta^\alpha
\sigma^\mu_{\alpha\dot\alpha} \bar\theta^{\dot \alpha},\;
\theta^\alpha,\;\bar\theta^{\dot\alpha},$$ with superfields
$$\Phi(x,\theta,\bar\theta)=\Phi''(\bar y,\theta,\bar\theta).$$ The brackets $\{\;,\;\}_{1}$ and $\{\;,\;\}_{4}$
would acquire simpler forms. The procedure is identical and we
will not repeat it here.

\subsection{ Moyal-Weyl star products}

Generically we consider Poisson brackets of the form
\begin{equation}\{\Phi, \Psi\}=\Phi\overleftarrow{\mathcal{D}_A}
P^{AB}\overrightarrow{\mathcal{D}_B}
\Psi,\label{pbg}\end{equation} where the index $A$ runs over all
the variables, even and odd. The derivations
$\overrightarrow{\mathcal{D}_A}$ commute (or anticommute) with
each other and $\overrightarrow{\mathcal{D}_C}P^{AB}=0$. The
matrix $P^{AB}$  has the right symmetry properties for (\ref{pbg})
to be a Poisson bracket. Under these assumptions there is an
associative star product of Moyal-Weyl type, defined by
\begin{equation}
\Phi\star \Psi= e^{hP}(\Phi, \Psi)=\sum_{n=0}^\infty
\frac{h^n}{n!}P^n(\Phi, \Psi)\label{stpd}
\end{equation}
where \begin{eqnarray*} &&P^n(\Phi, \Psi)=\sum_{A_1,\cdots
A_n;B_1,\cdots B_n}(-1)^{\rho_{A_1\cdots A_n}^{B_1\cdots
B_n}}\cdot\\&&
\Phi\overleftarrow{\mathcal{D}}_{A_1}\overleftarrow{\mathcal{D}}_{A_2}\cdots
\overleftarrow{\mathcal{D}}_{A_n} P^{A_1B_1}P^{A_2B_2}\cdots
P^{A_nB_n}\overrightarrow{\mathcal{D}}_{B_n}\cdots
\overrightarrow{\mathcal{D}}_{B_2}
\overrightarrow{\mathcal{D}}_{B_1} \Psi,\end{eqnarray*}
$$\hbox{and}\qquad\qquad \rho_{A_1\cdots A_n}^{B_1\cdots B_n}=
\sum_{i=1}^{n-1}(p_{A_i}+p_{B_i})\sum_{j=i+1}^np_{A_j}.$$ The
associativity of the star product
$$\phi\star (\psi\star\chi)=(\phi\star \psi)\star\chi,$$ follows
from the associativity in the purely even case (see for example
\cite{bffls}).

The sign $\rho$ it is needed to take into account the odd
character of $P^{\mu\alpha}$. The procedure is as follows. We
decompose again $A=(\mu, \alpha)$ and we define
$$\overrightarrow{\mathcal{K}}_a=\delta_{\mu
a}P^{\mu\alpha}\overrightarrow{\mathcal{D}}_\alpha,$$ where $a$
and $\mu$ run over the same set of numbers. Then
$\overrightarrow{\mathcal{K}}_a$ is an even derivation. The
Poisson bracket can be written as
$$\{\Phi, \Psi\}=P^{\mu\nu}
\Phi\overleftarrow{\mathcal{D}}_\mu
\overrightarrow{\mathcal{D}}_\nu \Psi+
P^{\alpha\beta}\Phi\overleftarrow{\mathcal{D}}_\alpha\overrightarrow{\mathcal{D}}_\beta\Psi
+ \delta^{\mu a }\bigl(\Phi\overleftarrow{\mathcal{D}}_\mu
\overrightarrow{\mathcal{K}}_a\Psi-
\Phi\overleftarrow{\mathcal{K}}_a \overrightarrow{\mathcal{D}}_\mu
\Psi\bigr). $$ This is a Poisson bracket with $\tilde P^{AB}$ of
block diagonal type (the index $A$ now runs over
$(\mu,a,\alpha)$). Then we can apply the standard formula for the
Moyal-Weyl product (see for example Ref. \cite{fl}) and obtain the
quantization. Returning to the previous notation gives the sign
$\rho$. In the case where $P^{AB}$ is block diagonal (that is,
$P^{AB}$ is always even), then  $\rho=0$.

 Notice that
because of
$$\frac {\partial\Phi}{\partial x^\mu}=\frac {\partial\Phi'}{\partial
y^\mu}=\frac {\partial\Phi''}{\partial \bar y^\mu},$$ the star
products defined by (\ref{stpd}) in each of the three sets of
coordinates $(x,\theta,\bar\theta)$, $(y,\theta,\bar\theta)$,
$(\bar y,\theta,\bar\theta)$ is exactly the same.

\bigskip

From the explicit formula for the quantization of the Poisson
bracket (\ref{pbg}), one has that if a derivation $\mathcal{K}$
(anti)commutes with $\mathcal{D}$,  it is also a derivation
either, of the Poisson bracket and of the corresponding star
product,
$$\mathcal{K}(\Phi\star\Psi)= (\mathcal{K}\Phi)\star
\Psi+(-1)^{p_\mathcal{K}p_\Phi}\Phi\star\mathcal{K}(\Psi).$$ So a
supersymmetric Poisson bracket gives rise to a supersymmetric star
product.

\bigskip

We remark again that in minkowskian signature the star products
 defined here do not satisfy a reality condition
$$\bar\Phi\star\bar\Psi\neq \overline{\Psi\star\Phi},$$
but they are perfectly consistent (for reality conditions on star
products, see for example Ref. \cite{de}). As for the Poisson
brackets we have
$$\bar\Phi\star_1\bar\Psi= \overline{\Psi\star_2\Phi},\qquad \bar\Phi\star_3\bar\Psi= \overline{\Psi\star_4\Phi}.$$

\section{\label{extended}Deformed extended superspace: chiral and harmonic superfields}

We consider now the case of a superspace with extended
supersymmetry, in which more  star products exist. In particular,
there exists a star product  that is not only invariant under
supertranslations but also Lorentz invariant. For definiteness we
consider a Poisson bracket of type $\{\;,\;\}_1$ with
$P^{\mu\nu}=P^{\mu\alpha}=0$. The same arguments could be used for
brackets of type $\{\;,\;\}_{2,3,4}$.

Let $i,j=1,\dots N$. We have \begin{equation}\{\Phi, \Psi\}=
P^{i\alpha \, j\beta  }\Phi\overleftarrow{D}_{\alpha
i}\overrightarrow{D}_{\beta j}\Psi.\label{extpb}\end{equation} The
matrix $P^{i\alpha \,j\beta }$ must be symmetric under the
simultaneous exchange $i\leftrightarrow j$ and
$\alpha\leftrightarrow\beta$. We can write
\begin{equation}P^{i\alpha \,j\beta }=P_s^{ij\,\alpha\beta }+P_a^{
i\, j}\epsilon^{\alpha \,\beta },\label{extenpb}\end{equation}
where the subscripts $a$ and $s$ mean that the matrix is
antisymmetric or symmetric respectively. The first term is symmetric under the independent
 exchanges $i\leftrightarrow j$ and
$\alpha\leftrightarrow\beta$.

 The second term is Lorentz invariant. Any Poisson
bracket containing only the second term will be Lorentz invariant,
and then super Poincar\'e invariant. This term has no analogue in
$N=1$ .

We could   choose $P_s^{ij\,  \alpha \beta}=\delta^{ij}P^{\alpha\beta}_s$. Then first term would be
$\rO(N)$ invariant.

\bigskip

For $N=2$ we could choose \begin{equation*}P^{\alpha i\,\beta j}=
P\epsilon^{ij}\epsilon^{\alpha\beta}.\end{equation*} Then, since
$\epsilon^{ij}$ is an invariant form of $\rSO(2)$ we can have
$\rSO(2)$ invariance and Lorentz invariance simultaneously. More
generally, we can decompose (\ref{extenpb}) in terms of
representations of $\rSO(2)$
$$P^{i\alpha \,j\beta }=\delta^{ij}P_{0s}^{\alpha \beta }+ P_{Ts}^{ij\,\alpha \beta }+
P_{0a}\epsilon^{ij}\epsilon^{\alpha \,\beta },$$ where the first
and last terms are $\rSO(2)$ singlets and the second term is the
$\rSO(2)$ doublet corresponding to the traceless part,
($P_{Ts}^{ij\,\alpha \beta }\delta_{ij}=0$). The second term does
not exist in $N=1$.

It is convenient to make the change of variables
$$
\theta^{\pm\alpha}=\frac 1{\sqrt{2}}(\theta^{1\alpha}\pm
i\theta^{2\alpha}), \qquad \bar\theta^{\pm\dot\alpha}=\frac
1{\sqrt{2}}(\theta^{1\dot\alpha}\pm i\theta^{2\dot\alpha})$$ so
that $(\theta^{+\alpha})^*=\bar\theta^{-\dot\alpha}$. The labels
$\pm$ are charges under $\rSO(2)$. (\ref{extenpb}) becomes
$$P^{\pm\alpha\;\mp\beta}=i(P_{0s}^{\alpha\beta}\mp
P_{0a}\epsilon^{\alpha\beta}),\qquad
 P^{\pm\alpha\;\pm\beta}=i(P_{Ts}^{1\alpha\, 1\beta}\pm P_{Ts}^{1\alpha\, 2\beta}).$$

This is a suitable basis to describe harmonic superspace
\cite{gios}. We make a shift in the superspace variables
$$x^\mu,\, \theta^\pm,\,\bar\theta^\pm\longrightarrow y^\mu=x^\mu+i\theta^+
\sigma^\mu\bar\theta^--i\theta^-\sigma^\mu\bar\theta^+,\,
\theta^\pm,\,\bar\theta^\pm.$$ In this basis the covariant
derivatives become
\begin{eqnarray*}
D^+_\alpha=\frac\partial{\partial\theta^{-\alpha}},\qquad
D^-_\alpha=-\frac\partial{\partial\theta^{+\alpha}}+
2i\bar\theta^{-\dot\alpha}\sigma^\mu_{\alpha\dot\alpha}\partial_\mu\\
\bar
D^+_{\dot\alpha}=\frac\partial{\partial\bar\theta^{-\dot\alpha}},\qquad
 \bar
D^-_{\dot\alpha}=-\frac\partial{\partial\bar\theta^{+\dot\alpha}}-
2i\theta^{-\alpha}\sigma^\mu_{\alpha\dot\alpha}\partial_\mu.\end{eqnarray*}

The change of basis can be expressed as
$$D_\alpha^\pm=U_i^\pm D^i_\alpha,$$
with $U_i^\pm$ being harmonics in $S^2$. The
 only non zero anticommutators are
$$\{D^+_\alpha,\bar D^-_{\dot\alpha}\}=-\{D^-_\alpha,\bar
D^+_{\dot\alpha}\}=-2i\sigma^\mu_{\alpha\dot\alpha}\partial_\mu.$$
 Chiral superfields satisfy
$$\bar D^\pm_{\dot\alpha}\Phi_c=0,$$
while harmonic superfields are defined as
$$ D^-_{\alpha}\Phi_h=\bar D^-_{\dot\alpha}\Phi_h=0.$$
In N=2 superspace, chiral superfields describe vector multiplets
while harmonic superfields describe hypermultiplets.

One could choose for example a deformation like,
\begin{eqnarray}\{\Phi, \Psi\}&=& P^{-\alpha,-\beta}
\Phi\overleftarrow{D}^+_{\alpha}\overrightarrow{ D}^+_{\beta
 }\Psi+P^{+\alpha,+\beta}
\Phi\overleftarrow{D}^-_{\alpha}\overrightarrow{ D}^-_{\beta
 }\Psi+\nonumber\\
&&+P^{+\alpha,-\beta}
\Phi\overleftarrow{D}^-_{\alpha}\overrightarrow{ D}^+_{\beta
 }\Psi+P^{-\alpha,+\beta}
\Phi\overleftarrow{D}^+_{\alpha}\overrightarrow{ D}^-_{\beta
 }\Psi.\label{pb1}\end{eqnarray} This has no analogue in $N=1$,
 if $P^{-\alpha,-\beta}$ or $P^{+\alpha,+\beta}$ is different from zero. Nevertheless,
  we have as before that antichiral superfields
have Poisson bracket zero, so their product is not deformed.

 Another choice could be  a deformation like \begin{eqnarray}\{\Phi,
\Psi\}&=& P^{-\alpha,-\dot\alpha}\bigr(
\Phi\overleftarrow{D}^+_{\alpha}\overrightarrow{\bar
D}^+{\dot\alpha }\Psi+ \Phi\overleftarrow{\bar
D}^+{\dot\alpha}\overrightarrow{D}^+_{\alpha }\Psi
\bigl)\nonumber\\&&+ P^{-\alpha,-\alpha}
\Phi\overleftarrow{D}^+_{\alpha}\overrightarrow{D}^+{\alpha }\Psi+
P^{-\dot\alpha,-\dot\alpha} \Phi\overleftarrow{\bar
D}^+{\dot\alpha}\overrightarrow{\bar D}^+_{\dot\alpha }\Psi
.\label{pb2}\end{eqnarray} Again, it has no analogue in $N=1$. In
this case the product in the subspace of {\it antiharmonic}
superfields ($D^+_{\alpha}\Phi_{ah}=\bar
D^+_{\dot\alpha}\Phi_{ah}=0$) is not deformed.

 If only one term,
say $P^{-\alpha,-\dot\alpha}$, is different from zero, then the
deformation is given in terms of the vector
 $v^\mu=P^{-\alpha,-\dot\alpha}\sigma^\mu_{\alpha\dot\alpha}.$

 As in section \ref{nonsusy}, we could do the same analysis with
 the operators $Q$'s instead of the operators $D$'s.

 In section \ref{commn=2} we will comment about the
physical consequences of these deformations.

\section{\label{wz}Wess-Zumino model in non commutative superspace}

In this section we want to explore the possibility of using the
star product to construct Lagrangians of physical theories. The
``purely even" deformation (that is, a deformation that is non
trivial on the coordinates $x^\mu$) has been studied in \cite{fl}.
In order to see the effects of the non commutativity of the odd
variables, we will set $P^{\mu\nu}=0$. We will analyze the theory
in both, euclidean and minkowskian signatures.

In the first place,  we will consider a Poisson bracket of type
$\{\;,\;\}_1$ and its quantization, which is supersymmetric. A
bracket of type $\{\;,\;\}_3$ was used in Ref.\cite{se}, with
which we will compare our results.

From now on, unless explicitly stated the star product
$\Phi\star\Psi$, without any subindex, will refer to the
Moyal-Weyl quantization of $\{\;,\;\}_1$ with
$P^{\mu\nu}=P^{\mu\alpha}=0$, so
\begin{equation*}\{\Phi, \Psi\}=
P^{\alpha\beta}\Phi\overleftarrow{D}_\alpha\overrightarrow{D}_\beta\Psi.\end{equation*}
The star product has a finite expansion \begin{eqnarray}\Phi\star
\Psi=\Phi\Psi +
hP^{\alpha\beta}\Phi\overleftarrow{D}_\alpha\overrightarrow{D}_\beta\Psi+
\frac {h^2} 4 \det P\Phi
\overleftarrow{D}^2\overrightarrow{D}^2\Psi=\nonumber\\= \Phi\Psi
+
hP^{\alpha\beta}(-1)^{(p_\Phi+1)}\overrightarrow{D}_\alpha\Phi\overrightarrow{D}_\beta\Psi-
\frac {h^2} 4 \det P\overrightarrow{D}^2\Phi
\overrightarrow{D}^2\Psi,\label{exp}\end{eqnarray} where
$$\overleftarrow{D}^2=\overleftarrow{D}_\alpha\overleftarrow{D}_\beta \epsilon^{\alpha\beta},
\qquad
\overrightarrow{D}^2=\epsilon^{\beta\alpha}\overrightarrow{D}_\alpha\overrightarrow{D}_\beta.$$
Another convenient way of expressing the star product is
\begin{equation}\Phi\star \Psi=\Phi\Psi + \overrightarrow{D}_{\alpha}\bigr(h
P^{\alpha\beta}(-1)^{(p_\Phi+1)}\Phi\overrightarrow{D}_{\beta}\Psi+
\frac {h^2} 4 \det P
\overrightarrow{D}^{\alpha}\Phi\overrightarrow{D}^2\Psi\bigl),\label{tld}\end{equation}
which makes manifest the fact that the  difference between the
ordinary product and the star product is a total covariant
derivative.

\bigskip

From now on we will use only left derivatives and we will denote
them simply as $D_\alpha, \bar D_{\dot \alpha}$. We will also
consider only even superfields, so $p_\Phi=0$,

The subalgebra of antichiral fields ($D\Phi_a=0$) does not get
deformed,
$$\Phi_a\star\Psi_a=\Phi_a\Psi_a.$$
More generally, the star product of an antichiral field with a
general field is the commutative product,
$$\Phi_a\star\Psi=\Phi_a\Psi.$$

\subsection{ \label{nssa}1/2 supersymmetric lagrangian in euclidean superspace}

In this subsection we consider a superspace with euclidean
signature, so $\theta$ and $\bar\theta$ are pseudoreal and
independent. We will see that the difficulty in defining a star
product that is simultaneously supersymmetric and has subalgebras
of chiral and antichiral superfields will lead to an explicit
breaking of supersymmetry in the Lagrangian.

Let $\Phi$ be a chiral superfield, and  $\bar \Phi$ an antichiral
superfield. The prescription for the kinetic term in the
Wess-Zumino lagrangian is
 $$\int\!\! d^2\!\theta \!d^2\!\bar\theta\,\Phi\star \bar \Phi=D^2\bar D^2\Phi\star \bar
 \Phi|_{\theta=\bar\theta=0}$$
which does not get deformed. The interaction terms that  involve
powers $\bar\Phi^{(\star n)}=\bar\Phi^{n}$ (in an obvious
notation) do not get deformed neither. These superfields are
antichiral, so the classical prescription for the Lagrangian
$$\int\!d^2\!\bar\theta\,\bar\Phi^{n}=\bar D^2\Phi\star \bar
 \Phi|_{\bar\theta=0}$$
 does not break  supersymmetry.

 We want to analyze the terms that can have a non
trivial contribution from the star product, as for example,
$\Phi^{(\star n)}$. This term is not a chiral superfield. But we
can still give the prescription
$$\int\!d^2\!\theta\,\Phi^{\star n}=D^2\Phi^{(\star
n)}|_{\theta=\bar\theta=0}.$$ We will see that this prescription
breaks $1/2$ of the supersymmetries, the $\bar Q_\alpha$, but it
is still invariant under the $Q_\alpha$.

 For $n=2$, and using
(\ref{tld}), we have
$$D^2(\Phi\star\Phi)=D^2(\Phi^2).$$ For $n=3$, and using (\ref{exp}) we have
\begin{equation}D^2(\Phi\star\Phi\star\Phi)=D^2(\Phi^3)-\frac{h^2}{4}\det
PD^2\Phi D^2\Phi D^2\Phi.\label{d2phi3}\end{equation} Let us
express the chiral field in terms of ordinary fields as usual
$$\Phi(y,\theta)=A(y)+ \theta\psi(y) +\theta\theta F(y).$$
Then
$$D^2(\Phi\star\Phi\star\Phi)|_{\theta=\bar\theta=0}=D^2(\Phi^3)|_{\theta=\bar\theta=0}-\frac{h^2}{4}\det
PF^3.$$ The term that is added to the action is proportional to
$F^3$.
 Since the term $\Phi^{(\star 3)}$ is non chiral, the Lagrangian
 cannot be invariant under the whole supersymmetry algebra. Let us
 see this statement in more detail. Let $\Psi$ be an antichiral
 superfield $D\Psi=0$,
 $$\Psi(\bar y, \bar\theta)=B(\bar y)+ \bar\theta\bar\chi(\bar y) +\bar\theta\bar\theta G(\bar
 y).$$The supersymmetry transformations are
 \begin{eqnarray}
 &\delta_\epsilon B=0\qquad &\delta_{\bar \epsilon}
 B= \bar\epsilon\bar \chi\nonumber\\
 &\delta_\epsilon\bar\chi=-i {2} \epsilon\sigma^\mu\partial_\mu B\qquad&\delta_{\bar
 \epsilon}\bar\chi= {2}G\bar\epsilon\nonumber\\
&\delta_\epsilon G=i\epsilon\sigma^\mu\partial_\mu\bar\chi\qquad
&\delta_{\bar \epsilon}
 G=0\label{susyt}\end{eqnarray}
 The
  component $B$ of an antichiral superfield is invariant under the supersymmetries $Q$.
  If $\bar\chi$ were a total derivative, then the integral of $B$ on space time
  would be
  invariant under both, the $Q$ and $\bar Q$ supersymmetries.
In particular $D^2\Theta$ is an antichiral
 superfield for arbitrary $\Theta$, and if $\Theta$ is itself a chiral
 superfield then the component $\bar\chi$ is a total derivative.

 In  this case,
 the lagrangian is constructed as $D^2\Theta$, but $\Theta$
 contains non chiral terms. As a consequence, the action can only
 be invariant under half of the supersymmetry generators.

 It is remarkable that the Lagrangian that we obtain is the same that
 the one
 proposed in Ref.  \cite{se}. There, the star product chosen was the star product
quantizing a bracket of type $\{\;,\;\}_3$ in (\ref{suse}). This
star product breaks the $\bar Q$ supersymmetries explicitly, and
although chiral fields are a good subalgebra of the star product,
the resulting Lagrangian preserves only half of the
supersymmetries.

\bigskip

It is easy to compute some higher order terms in the Lagrangian,
\begin{eqnarray*}D^2(\Phi^{(\star 4)})&=&D^2\Phi^4
-\frac 1 2 \det P(D^2\Phi^2)D^2\Phi D^2\Phi,\\\\
D^2(\Phi^{(\star 5)})&=&D^2\Phi^4 +\frac 1 2 \det
P(D^2\Phi)^2D^2\Phi^3 \\&&+\frac 1 4 \det
P(D^2\Phi^2)^2D^2\Phi+\frac 1 {16} (\det P)^2(D^2\Phi)^5.
\end{eqnarray*}

As mentioned in Ref. \cite{se}, the terms appearing in the
Lagrangian contain $P^{\alpha\beta}$ only through the expression
$\det P$, so the action is Lorentz invariant.

Quantum properties of the model with the  deformation involving the $Q_\alpha$ operators  have been studied in
Rfs. \cite{bfr,ty}

\subsection{\label{commn=2}A comment on N=2 theories.}
We remind that for $N=2$ theories we considered  two types of
deformations, (\ref{pb1}) and (\ref{pb2}). One could construct
theories with the method we have used for $N=1$. In the first case
(the chiral case), one would expect that the $\bar
Q^\pm_{\dot\alpha}$  supersymmetry will be broken, while in the
second case (the harmonic case) the broken supersymmetries will be
$\bar Q^+_{\dot\alpha}, Q^+_{\alpha}$, .

Since $N=2$ theories may contain both, chiral and harmonic
superfields, such a theory will have three broken supersymmetries,
$\bar Q^\pm_{\dot\alpha}$ and $\bar Q^+_{\dot\alpha}$. This is in
agreement with Ref. \cite{bs}.

\subsection{An example with $P^{\mu\alpha}\neq 0$.}
In this section we want to consider  an example of a deformation
with $P^{\mu\alpha}\neq 0$. Generically $P^{\mu\alpha}\neq 0$
contains $4\times 2$  odd parameters, so the expansion of the star
product (\ref{stpd}) will necessarily end at order 8. We consider
here the following ansatz: We take $P^{\mu\alpha}=0$ for $\mu\neq
1$ and $P^{1\alpha}\neq 0$. We denote $\partial_1=\partial$ and
$P^2=P^{1\alpha}P_{1\alpha}$. We have
\begin{eqnarray*}\Phi\star\Phi&=&\Phi^2+\frac
{h^2}2P^2(D^\alpha\partial\Phi
D_\alpha\partial\Phi-D^2\Phi\partial^2\Phi)=\\&& \Phi^2+\frac
{h^2}2P^2\bigl((D^\alpha(\partial\Phi\partial
D_\alpha\Phi)-\partial(\partial\Phi
D^2\Phi)\bigr)=\\&&\Phi^2+\frac
{h^2}2P^2\bigl(\frac12D^2(\partial\Phi)^2-\partial(\partial\Phi
D^2\Phi)\bigr).\end{eqnarray*}

We want to compute the contribution of $\Phi\star\Phi\star\Phi$ to
the Lagrangian,
\begin{eqnarray*}D^2\bigl((\Phi\star\Phi)\Phi\bigl)&=& D^2(\Phi^3 +\frac
{h^2}2P^2\bigl(\frac12D^2(\partial\Phi)^2\Phi-\partial(\partial\Phi
D^2\Phi)\Phi\bigr)\bigr)\backsimeq\\&&D^2\Phi^3 +\frac 3 2\frac
{h^2}2D^2(\partial\Phi)^2D^2\Phi^2,
\end{eqnarray*}
modulo terms that are total spacetime derivatives. The deformation
term is proportional to
\begin{eqnarray*}P^2F(\frac 14 \partial\psi\partial\psi-\partial A\partial F)\backsimeq
P^2\bigr(\frac 14(\partial\psi^\alpha\partial\psi_\alpha)F+\frac
12F^2\Box A\bigr)
\end{eqnarray*}

\bigskip

The term of order 1 in $h$ will appear only if we take more than
one superfield. We could have, for example
$(\Phi_1\star\Phi_2)\Phi$. Assume for simplicity that only one
Grassmann parameter $P^{\mu\alpha}$ is different from zero, so
only the first order contributes. We keep nevertheless the Lorentz
covariant notation. Then, up to total spacetime derivatives
\begin{eqnarray*}D^2\bigl((\Phi_1\star\Phi_2)\Phi\bigr)&\backsimeq&
P^{\mu\alpha}\bigl(\Phi_1D_\alpha\Phi_2\partial_\mu\Phi-\Phi_1\partial_\mu\Phi_2D_\alpha\Phi\bigr)\backsimeq\\
&&P^{\mu\alpha}\bigl(((A_1\psi_{2\alpha})-(A_2\psi_{1\alpha}))\partial_\mu
F-(A_1F_2-A_2F_1)\partial_\mu\psi_\alpha\\&&
+F(\psi_{1\alpha}\partial_\mu
A_2-\psi_{2\alpha}\partial_\mu A_1+A_1\partial_\mu\psi_{2\alpha}-A_2\partial_\mu\psi_{1\alpha})-\\
&&(F_1\partial_\mu A_2-F_2\partial_\mu A_1+A_1\partial_\mu
F_2-A_2\partial_\mu F_1)\psi_\alpha\bigr).\end{eqnarray*}

\subsection{ \label{ssdwzm} Supersymmetric lagrangian in minkowskian superspace}
We consider now a superspace with minkowskian signature. We remind
that the star product does not have a reality condition,
\begin{equation}\bar\Phi\star\bar\Psi\neq \overline{\Psi\star\Phi}.\label{nrc}\end{equation} $\Phi$
and $\bar\Phi$ are now related by complex conjugation. The kinetic
term $\Phi\star\bar\Phi=\Phi\bar \Phi$ is real and not deformed.
We consider terms of two types, with the following prescriptions:

\smallskip

\noindent 1. $\Phi^{(\star n)}$ is a non-chiral superfield. It
contributes to the action as
$$\int\!\! d^2\!\theta \!d^2\!\bar\theta\,\Phi^{(\star n)}=D^2\bar D^2\Phi^{(\star n)}|_{\theta=\bar\theta=0}.$$
This becomes zero in the classical limit, $h\rightarrow 0$, so it
is a purely non commutative correction. It is not real, so we
follow the standard prescription of adding the hermitian conjugate
term to the action,
$$\int\!\! d^2\!\theta \!d^2\!\bar\theta\,\overline{\bigl(\Phi^{(\star n)}\bigr)}=
D^2\bar D^2\overline{\bigl(\Phi^{(\star
n)}\bigr)}|_{\theta=\bar\theta=0}.$$ Notice that because of
(\ref{nrc}), $\overline{\bigl(\Phi^{(\star n)}\bigr)}\neq
\bar\Phi^{(\star n)}$.

\smallskip

\noindent 2. $\bar \Phi^{(\star n)}= \bar \Phi^{n}$ is an
antichiral superfield. We introduce it in the action as
$$\int\!\!d^2\!\bar\theta\,\bar\Phi^{n}=\bar D^2\bar \Phi^{n}|_{\bar\theta=0}.$$
Again, we should  add the hermitian conjugate term
$$\int\!\!d^2\!\theta\,\overline{(\bar\Phi^{n})}= D^2\overline{(\bar\Phi^{n})}|_{\theta=0}=
D^2{(\Phi^{n})}|_{\theta=0}.$$

The kinetic term plus  terms of the type 1 and 2 (together with
their hermitian conjugates) give a deformed, supersymmetric
Wess-Zumino model, formulated in a non commutative superspace with
a non trivial deformation of the odd variables, and that has the
correct classical limit.

The deformation of  $\Phi^{(\star 2)}=\Phi^2$  is a total
$(\partial, D)$ derivative.

Notice that one can use either the star product corresponding to a
Poisson bracket of type $\{\;,\;\}_1$ (involving the $D$'s), or a
star product associated with the Poisson bracket $\{\;,\;\}_2$
(involving the $\bar D$'s). The prescription to compute the
Lagrangian will lead to the same Lagrangian. So both star products
can be used interchangeably in the action, and have the same
physical meaning.

\bigskip
Let us compute the term of type 1. For $n=3$. We have (using
(\ref{d2phi3})) \begin{eqnarray*}&&\bar D^2
D^2(\Phi\star\Phi\star\Phi)|_{\theta=\bar\theta=0}= \\&&=\bar
D^2(D^2(\Phi^3)-\frac{h^2}{4}\det PD^2\Phi D^2\Phi
D^2\Phi)|_{\theta=\bar\theta=0}=\\&&=\frac{h^2}{4}\det P
\bigr(\frac
{1}{2}\sigma_{\alpha\dot\alpha}^\mu\sigma_{\beta\dot\beta}^\nu\epsilon^{\dot
\alpha\dot \beta}
\partial_\mu\psi^{\alpha}\partial_\nu\psi^{\beta}
F+F^2\Box A\bigl).\end{eqnarray*}

Notice that the correction to the Wess-Zumino action computed in
Section \ref{nssa} corresponds to the first component of the
chiral superfield $(D^2\Phi)^3$, and consequently breaks 1/2
supersymmetry ($\bar Q$) (see (\ref{susyt})). In minkowskian
spacetime, adding the hermitian conjugate of this term will break
the other half of the supersymmetry ($Q$), so the resulting action
will not be invariant under any supersymmetry.

Instead, the correction that we compute here is the last component
of the same superfield $(D^2\Phi)^3$. The variation of the last
component of a chiral superfield under supersymmetry
transformations is a total spacetime derivative, and the action is
supersymmetric.

\section*{Acknowledgments}

S. F. and O. M. want to thank the INFN, Sezione di Torino and the
Politecnico di Torino for their hospitality during the realization
of this work.

M. A. Ll. wants to thank the Departament de F\'{\i}sica Te\`orica
of the Universitat de Val\`encia and the IFIC for their
hospitality during the realization of this work.

O. M. wants to thank the  the Ministerio de Ciencia y
Tecnolog\'{\i}a, Spain, for his FPI research grant.

 Work supported in part by  the European
Community's Human Potential Program under contract
HPRN-CT-2000-00131 Quantum Space-Time, in which M. A. Ll. is
associated to Torino University.

The work of S. F. has also  been supported by the D.O.E. grant
DE-FG03-91ER40662, Task C.

The work of M. A. Ll. and O. M. has also been supported by the
research grant BFM 2002-03681 from the Ministerio de Ciencia y
Tecnolog\'{\i}a (Spain) and from EU FEDER funds.

\end{document}